\documentstyle[aps,12pt]{revtex}

\begin{document}
\author{E.N. Glass\thanks{%
Permanent address: Physics Department, University of Windsor, Ontario N9B
3P4, Canada}}
\address{Department of Physics, University of Michigan, Ann Arbor, Michgan 48109}
\date{31 July 2001, \ \ \ preprint MCTP-01-35}
\title{A CONSERVED BACH CURRENT}
\maketitle

\begin{abstract}
\ \newline
The Bach tensor and a vector which generates conformal symmetries allow a
conserved four-current to be defined. The Bach four-current gives rise to a
quasilocal two-surface expression for power per luminosity distance in the
Vaidya exterior of collapsing fluid interiors. This is interpreted in terms
of entropy generation. \newline
\newline
PACS numbers: 04.20.Jb, 04.40.Nr\newpage\
\end{abstract}

\section{INTRODUCTION}

Gravitational theory allows a conserved current $J_{\text{{\sc EIN}}%
}^{\alpha }$ to be formed from the Einstein tensor with a symmetry generated
by Killing vector $k^{\beta }$:
\[
J\text{$_{\text{{\sc EIN}}}^{\alpha }$}=G_{\ \beta }^{\alpha }k^{\beta }.
\]
The current $J_{\text{{\sc EIN}}}^{\alpha }$ is conserved since the Einstein
tensor is divergence-free and $k^{\beta }$ satisfies Killing's equation. In
a similar fashion we define
\begin{equation}
J_{\text{{\sc BACH}}}^{\alpha }:=B_{\ \beta }^{\alpha }\xi ^{\beta }
\label{bach-cur}
\end{equation}
where $B_{\alpha \beta }$ is the Bach tensor, a symmetric, divergence-free,
trace-free tensor \cite{Sze68},\cite{KNT85},\cite{PR86v2},\cite{Que98}. $\xi
^{\beta }$ is a generator of conformal maps which include Killing symmetry
and homothety as special cases. $J_{\text{{\sc BACH}}}^{\alpha }$ is
conserved because the Bach tensor is divergence-free and the symmetrized
covariant derivative of the conformal generator $\xi ^{\beta }$ is\ $\xi
_{(\alpha ;\beta )}=2\Psi (x^{\nu })g_{\alpha \beta }$. Since the Bach
tensor is trace-free it follows that $B_{\alpha \beta }\xi ^{(\alpha ;\beta
)}$ vanishes.

The Bach tensor is a second derivative of the Weyl tensor $C_{\alpha \mu
\beta \nu }$ plus a Ricci tensor term
\begin{equation}
B_{\alpha \beta }=\nabla ^{\mu }\nabla ^{\nu }C_{\alpha \mu \beta \nu }+%
\frac{1}{2}R^{\mu \nu }C_{\alpha \mu \beta \nu }.  \label{bach1}
\end{equation}
For metrics with non-zero Bach tensor, we can think of $B_{\alpha \beta
}(g)=\kappa _{weyl}S_{\alpha \beta }$ as the field equations of conformal
Weyl gravity \cite{DS00}, with symmetric $S_{\alpha \beta }$ analogous to $%
T_{\alpha \beta }$ of Einstein's theory.

There exist superpotentials for $J_{\text{{\sc EIN}}}^{\alpha }$ which lead
to an Einstein mass by integrating the superpotential over two-surface $%
S^{2} $. Three of the superpotentials do not require a coordinate choice:
the Komar potential \cite{Kom59}, the Penrose-Goldberg potential \cite{Gol90}%
, and the Taub potential \cite{Gla93}. However, each of those three
superpotentials is restricted to special systems. For more general
potentials derived from the Hilbert action and pseudotensors, the Einstein
energy is known to be ambiguously defined because it depends on the choice
of background coordinates. Nevertheless, one may use the two-surface
construction to define sensible notions of total gravitational energy.
Indeed, for asymptotic flatness, say, towards future null infinity ${\cal I}%
^{+}$, $S^{2}$ tends to a ''nice'' cut of ${\cal I}^{+}$, a round
two-sphere. The (suitably unique) choice of asymptotically Cartesean
coordinates ensures that the Einstein energy agrees with the accepted
Bondi-Sachs notion of total energy \cite{Gol80} within some quasilocal (i.e.
finite) two-surface $S^{2}$. One of the first fundamental expositions of
quasilocal mass (and angular momentum) was given by Penrose \cite{Pen82}. A
more recent and well known treatment of quasilocal energy is the work of
Brown et al \cite{BLY99}.

In this work we introduce the conserved Bach current as a tool, in general
relativity, to study collapsing star models. In particular, we use the
properties of the Bach tensor and Bach current to study collapsing spherical
fluid interiors matched to a Vaidya exterior and analyze the power flow. $%
\xi ^{\alpha }$ generates a {\it matter symmetry} in the Vaidya exterior,
weaker than conformal symmetry, but $\xi _{(\alpha ;\beta )}$ annihilates
the Bach tensor and so the current is conserved. The volume integral of $J_{%
\text{{\sc BACH}}}^{\alpha }$ leads to a two-surface integral which provides
a quasilocal value for the Vaidya power gradient. Covariant thermodynamics
of irreversible systems is used to interpret the Bach current for the Vaidya
exterior. The Vaidya power gradient is related to the rate of entropy
increase and the number of states the energy resides in.

We can view a star model as a single state drawn from a microcanonical
ensemble. The initial values of all the observables are at their equilibrium
values (say, all zero), then, as a function of time, they will not remain
exactly zero but will execute small fluctuations about their zero
equilibrium values. The Vaidya exterior dissipates energy by sending
''short-wavelength photons'' to null infinity. It is the Bach current and
the rate of entropy increase that allows one, via the
fluctuation-dissipation theorem, to measure the rate at which a deviation
from equilibrium decays.

\ \newpage

{\bf CONVENTIONS AND NOTATION}

In this work Greek indices range over (0,1,2,3) = ($u,r,\vartheta ,\varphi $%
). Sign conventions are $2A_{\nu ;[\alpha \beta ]}=A_{\mu }R_{\ \nu \alpha
\beta }^{\mu },$ and $R_{\alpha \beta }=R_{\ \alpha \beta \nu }^{\nu }.$ $\ $%
Overdots abbreviate $\partial /\partial u$ and primes abbreviate $\partial
/\partial r$. The metric signature is (+,-,-,-) and the field equations are $%
G_{\mu \nu }=-8\pi T_{\mu \nu }$.

\section{SOME PROPERTIES OF THE BACH TENSOR}

\ \newline
1. A variational derivative of the Lagrangean density\ ${\cal L}=\sqrt{-g}%
C_{\alpha \beta \gamma \delta }C^{\alpha \beta \gamma \delta }$ \ yields the
Bach tensor
\begin{equation}
\sqrt{-g}B_{\alpha \beta }=\frac{\delta {\cal L}}{\delta g^{\alpha \beta }},
\label{bach-lagr}
\end{equation}
and the Lagrangean provides the field equation
\begin{equation}
B_{\alpha \beta }(g)=0.  \label{bach-zero}
\end{equation}
Solutions of Eq.(\ref{bach-zero}) are vacuum solutions of conformal Weyl
gravity \cite{DS00}.\newline
2. Kozameh et al \cite{KNT85} proved that any solution of Eq.(\ref{bach-zero}%
) which is also conformal to $C$ spaces is conformal to an Einstein metric ($%
\mathop{\rm Ei}%
$nstein metrics have Ricci tensor proportional to the metric). For an
example of a metric which satisfies $B_{\alpha \beta }(g)=0$ but is not an
Einstein metric see Nurowski and Plebanski \cite{NP00}.\newline
3. Vacuum spacetimes, conformally flat spacetimes, and Einstein spacetimes
have vanishing Bach tensor.\newline
4. For $P_{\mu \nu }=R_{\mu \nu }-(R/6)g_{\mu \nu }$, the Bianchi identities
allow the alternate expression
\begin{equation}
B_{\alpha \beta }=\nabla ^{\mu }\nabla _{\lbrack \alpha }P_{\mu ]\beta }+%
\frac{1}{2}P^{\mu \nu }C_{\alpha \mu \beta \nu }.  \label{bach2}
\end{equation}
5. Under conformal map $\hat{g}_{\alpha \beta }=\Omega ^{2}g_{\alpha \beta }$%
, the Bach tensor transforms as $\hat{B}_{\alpha \beta }=B_{\alpha \beta }$.

\section{THE VAIDYA SPACETIME}

\subsection{Vaidya Interiors}

The first Vaidya interior was constructed by Synge \cite{Syn57}. In that
work an incoherent shell of radiation is joined to Vaidya. Santos \cite
{San85} joined a collapsing fluid with radial heat flow to a Vaidya
exterior. Glass \cite{Gla90} studied a two-fluid interior, a collapsing
perfect fluid with an outgoing null fluid matched to a Vaidya exterior.
Continuity of the first and second fundamental forms expressed conservation
of momentum across the boundary and revealed the boundary to be the perfect
fluid zero-pressure surface. This is the interior we have in mind to model
the energy flow considered here. As the perfect fluid collapses, energy is
removed by the null fluid. In the Vaidya exterior, the outgoing
short-wavelength photons carry energy to future null infinity.

\subsection{Vaidya Exterior}

The Vaidya metric in outgoing null coordinates, with arbitrary mass function
$m(u)$ and $A:=1-2m(u)/r$, is given by
\begin{equation}
g_{\alpha \beta }^{Vad}dx^{\alpha }dx^{\beta
}=Adu^{2}+2dudr-r^{2}(d\vartheta ^{2}+\sin ^{2}\vartheta \ d\varphi ^{2}).
\label{vad_met}
\end{equation}
We use a Newman-Penrose null tetrad:
\begin{mathletters}
\begin{eqnarray}
l^{\alpha }\partial _{\alpha } &=&\partial _{r},\ \ \ \ \ \ \ \ \ \ \ \ \ \
\ \ \ \ \ \ \ \ \ \ \ \ \ \ \ \ \ \ l_{\alpha }dx^{\alpha }=du,
\label{tet_a} \\
n^{\alpha }\partial _{\alpha } &=&\partial _{u}-(A/2)\partial _{r},\ \ \ \ \
\ \ \ \ \ \ \ \ \ \ \ \ n_{\alpha }dx^{\alpha }=dr+(A/2)du,  \label{tet_b} \\
m^{\alpha }\partial _{\alpha } &=&(r^{-1}/\sqrt{2})(\partial _{\vartheta }+%
\frac{i}{\text{sin}\vartheta }\partial _{\varphi }),\ \ \ m_{\alpha
}dx^{\alpha }=-(r/\sqrt{2})(d\vartheta +i\text{sin}\vartheta \ d\varphi ),
\label{tet_c}
\end{eqnarray}
with non-zero spin coefficients
\end{mathletters}
\begin{mathletters}
\begin{eqnarray}
\rho &=&-1/r,\ \ \ \ \mu =-A/(2r), \\
\alpha &=&-\text{cot}\vartheta /(2\sqrt{2}r)=-\beta , \\
\gamma &=&A^{\prime }/4.
\end{eqnarray}
The Weyl tensor is given in terms of a basis set of anti self-dual bivectors
($\Psi _{2}=-m(u)/r^{3}$)
\end{mathletters}
\begin{equation}
C^{\alpha \mu \beta \nu }=[m(u)/r^{3}](U^{\alpha \mu }V^{\beta \nu
}+V^{\alpha \mu }U^{\beta \nu }+M^{\alpha \mu }M^{\beta \nu }+c.c.),
\label{vad_weyl}
\end{equation}
where
\begin{equation}
U^{\alpha \beta }=2\bar{m}^{[\alpha }n^{\beta ]},\text{ \ }M^{\alpha \beta
}=2l^{[\alpha }n^{\beta ]}-2m^{[\alpha }\bar{m}^{\beta ]},\text{ \ }%
V^{\alpha \beta }=2l^{[\alpha }m^{\beta ]}.  \label{asd-bivec}
\end{equation}
The Vaidya spacetime is Petrov type {\bf D} and has the invariant relation
\begin{equation}
R_{\alpha \beta \mu \nu }R^{\alpha \beta \mu \nu }=C_{\alpha \beta \mu \nu
}C^{\alpha \beta \mu \nu }=48\Psi _{2}^{2}.  \label{vad-inv}
\end{equation}
The Ricci and energy-momentum tensors are
\begin{equation}
R_{\mu \nu }=-8\pi T_{\mu \nu }=\left( \frac{2\dot{m}}{r^{2}}\right) l_{\mu
}l_{\nu }.  \label{vad_ric}
\end{equation}
The standard interpretation of the Vaidya energy flow was given by Lindquist
et al \cite{LSM65} with respect to an observer at rest at future null
infinity whose worldline tangent is
\begin{equation}
\hat{u}^{\mu }\partial _{\mu }=A^{-1/2}\partial _{u}=A^{-1/2}(n^{\mu }+\frac{%
1}{2}A\,l^{\mu })\partial _{\mu }.  \label{unit-v}
\end{equation}
The local energy flux $f$ is
\begin{equation}
T_{\mu \nu }\hat{u}^{\mu }\hat{u}^{\nu }=-\left( \frac{\dot{m}}{4\pi r^{2}}%
\right) \frac{1}{1-2m(u)/r},  \label{energy-flux}
\end{equation}
and the total luminosity seen at infinity is $L_{\infty }(u)=\
_{r\rightarrow \infty }^{\lim }\ 4\pi r^{2}f=-\dot{m}$.

Substitution of the Weyl tensor Eq.(\ref{vad_weyl}) and the Ricci tensor Eq.(%
\ref{vad_ric}) into Eq.(\ref{bach1}) yields the Bach tensor
\begin{equation}
B_{\alpha \beta }(g^{Vad})=(\frac{2\ddot{m}}{r^{3}})l_{\alpha }l_{\beta }+(%
\frac{2\dot{m}}{r^{4}})(l_{\alpha }n_{\beta }+n_{\alpha }l_{\beta
}+m_{\alpha }\bar{m}_{\beta }+\bar{m}_{\alpha }m_{\beta }).  \label{bach3}
\end{equation}
The independent eigenvectors of $B_{\alpha \beta }(g^{Vad})$ are $\partial
_{r}=l^{\alpha }\partial _{\alpha }$, $\partial _{\vartheta }=(r/\sqrt{2}%
)(m^{\alpha }+\bar{m}^{\alpha })\partial _{\alpha }$, and $\partial
_{\varphi }=-i(r/\sqrt{2})$sin$\vartheta (m^{\alpha }-\bar{m}^{\alpha
})\partial _{\alpha }$.

\section{VAIDYA CONSERVED CURRENT}

The generator of time translations is $\partial _{u}=\xi ^{\alpha }\partial
_{\alpha }$. The time dependence of $m(u)$ prevents $\xi ^{\alpha }$ from
being a Killing symmetry but $\xi ^{\alpha }$ generates a {\it matter
symmetry} since ${\cal L}_{\xi }g_{\alpha \beta }^{Vad}=-rR_{\alpha \beta }$%
. The Bach current $J_{\text{{\sc BACH}}}^{\alpha }=B^{\alpha \beta }\xi
_{\beta }$ for metric $g^{Vad}$ will be labelled $J_{Vad}^{\alpha }$. The
four-current is conserved since $B_{\alpha \beta }\xi ^{(\alpha ;\beta
)}=B_{\alpha \beta }l^{\alpha }l^{\beta }=0$. Stronger symmetry can be
imposed by restricting the mass function. With a linear mass function, the
Vaidya metric has been used to study self-similar collapse \cite{SV00}. This
restriction was anticipated by the author \cite{Gla99} for $%
m(u)=m_{0}+m_{1}u $ and homothetic symmetry $\xi ^{\alpha }\partial _{\alpha
}=(u_{0}+u)\partial _{u}+\partial _{r}$ where ${\cal L}_{\xi }g_{\alpha
\beta }^{Vad}=2g_{\alpha \beta }^{Vad}$.

With $\xi ^{\alpha }=(A/2)l^{\alpha }+n^{\alpha }$, the conserved
four-current is
\begin{equation}
J_{Vad}^{\alpha }=\left( \frac{2\dot{m}}{r^{4}}\right) n^{\alpha }+\left(
\frac{2\ddot{m}}{r^{3}}+A\frac{\dot{m}}{r^{4}}\right) l^{\alpha }.
\label{vad_j1}
\end{equation}
The four-current is more transparent when written in terms of curve tangents
\begin{equation}
J_{Vad}^{\alpha }\partial _{\alpha }=\left( \frac{2\dot{m}}{r^{4}}\right)
\partial _{u}+\left( \frac{2\ddot{m}}{r^{3}}\right) \partial _{r},
\label{vad_j2}
\end{equation}
and is similar to a conserved Maxwell current $\sqrt{-g}J_{Max}^{\alpha }=%
\sqrt{-g}(\rho ,\vec{J})$. By analogy we call $2\dot{m}/r^{2}$ the Bach
''charge density''. Examining $\sqrt{-g}J_{Vad}^{\alpha }$ tells one that $2%
\dot{m}/r^{2}$ is the flow across $u=const$ null three-surfaces and $2\ddot{m%
}/r$ is the flow across $r=const$ timelike three-surfaces. This would be
equally apparent upon integrating $\partial _{\alpha }(\sqrt{-g}%
J_{Vad}^{\alpha })=0$ over a four-volume bounded by four such surfaces.

A potential exists for $J_{Vad}^{\alpha }$, namely the bivector density
\begin{equation}
U_{Vad}^{\alpha \beta }=\sqrt{-g}(\frac{\dot{m}}{r^{3}})4l^{[\alpha
}n^{\beta ]}  \label{bivec-pot}
\end{equation}
with $\sqrt{-g}J_{Vad}^{\alpha }=\partial _{\beta }U_{Vad}^{\alpha \beta }.$
We integrate over a $u=const$ null three-surface ${\cal N}$:
\begin{equation}
\int\limits_{{\cal N}}\sqrt{-g}J_{Vad}^{\alpha }dS_{\alpha
}=\oint\limits_{\partial {\cal N}}U_{Vad}^{\alpha \beta }dS_{\alpha \beta }
\label{u-stokes}
\end{equation}
where $dS_{\alpha }=l_{\alpha }drd\vartheta d\varphi $ and $\partial {\cal N}
$ is a closed two-surface bounding ${\cal N}$ with $dS_{\alpha \beta
}=l_{[\alpha }n_{\beta ]}d\vartheta d\varphi $. The integral of the Bach
''charge density''\ over $\partial {\cal N}$ provides the quasilocal Bach
charge within $\partial {\cal N}$, which here is the Vaidya luminosity
divided by luminosity distance (and so we call it ''luminosity gradient'')
\begin{equation}
\oint\limits_{\partial {\cal N}}(\frac{\dot{m}}{r})\text{sin}\vartheta
\,4l^{[\alpha }n^{\beta ]}dS_{\alpha \beta }=-8\pi \frac{\dot{m}}{r},
\label{lum-grad}
\end{equation}
where $r$ ranges from the spacelike surface $r=2m$ to future null infinity.

\section{THERMODYNAMIC INTERPRETATION}

The Eckart theory of dissipative fluids is the simplest relativistic
generalization of Navier-Stokes theory, but Eckart's theory suffers from a
lack of stable solutions and acausal propagation of perturbations. We will
follow Calzetta \cite{Cal98} and use the framework of more causal and stable
theories such as the Israel-Stewart `second order' type theory or the
Geroch-Lindblom `divergence type' description of a relativistic real fluid.
A covariant theory has the following rules:

(a) Intensive quantities ($T,p,\mu $) are associated with scalars, which
represent the value of the quantity at a given event, as measured by an
observer at rest with respect to the fluid.

(b) Conjugate extensive quantities ($S,V,N$) are associated with vector
currents ($S^{\alpha },\hat{u}^{\alpha },N^{\alpha }$). If any of the
extensive currents $X^{\alpha }$ is conserved, then $X_{;\alpha }^{\alpha
}=0 $. The quantity $\hat{u}^{\alpha }$ associated with the volume is the
fluid unit four-velocity.

(c) Energy and momentum are combined into a single extensive quantity and
associated with the tensor $T^{\alpha \beta }$. The energy current is $%
U^{\alpha }=T^{\alpha \beta }\hat{u}_{\beta }$.

The entropy current $S^{\alpha }$ is given by
\begin{equation}
TS^{\alpha }=T^{\alpha \beta }\hat{u}_{\beta }+p\hat{u}^{\alpha }-\mu
N^{\alpha },  \label{ts-vec}
\end{equation}
which is rewritten as
\begin{equation}
S^{\alpha }=T^{\alpha \beta }\beta _{\beta }+\Phi ^{\alpha }-\alpha
N^{\alpha }  \label{s-vec}
\end{equation}
with affinity $\alpha =\mu /T$, thermodynamic potential $\Phi ^{\alpha
}=p\beta ^{\alpha }$, and inverse temperature vector $\beta ^{\alpha }=\hat{u%
}^{\alpha }/T$.\ $N^{\alpha }=n\hat{u}^{\alpha }$, where $n$ is the particle
number density seen by a comoving observer.

Suppose the fluid departs from equilibrium by a fluctuation $\delta
N^{\alpha }$, $\delta T^{\alpha \beta }$, consistent with the conservation
laws but otherwise arbitrary. Then the change in entropy production is
\begin{equation}
\delta S_{\,;\alpha }^{\alpha }=-\alpha _{,\alpha }\delta N^{\alpha }+\beta
_{\beta ;\alpha }\delta T^{\alpha \beta }.
\end{equation}
For a true equilibrium state the entropy must be stationary \cite{Isr87},
and so $\alpha _{,\alpha }=\beta _{\left( \beta ;\alpha \right) }=0$. Thus
the affinity must be constant, and the inverse temperature vector must be
Killing.

If $\xi ^{\mu }$ generates a timelike conformal symmetry then one has
thermal equilibrium with inverse temperature $\beta ^{\mu }=\xi ^{\mu }/T$
as a conformal Killing vector for trace-free $\delta T^{\alpha \beta }$. In
the exterior Vaidya region we have a weaker matter symmetry ${\cal L}_{\xi
}g_{\alpha \beta }^{Vad}=-rR_{\alpha \beta }$ and so $\delta S_{\,;\alpha
}^{\alpha }\neq 0$. The physical dimensions of $\dot{S}$ are energy/(sec$\
^{\circ }$K) and, reasoning dimensionally,
\begin{equation}
\frac{\dot{m}}{r}\left( \frac{k_{B}G}{c^{4}}\right) \sim \dot{S}
\end{equation}
in units with $k_{B}=G=c=1$. If one examines the terms of Eq.(\ref{s-vec})
its clear that $\Phi ^{\alpha }$ and $N^{\alpha }$ lie along $\hat{u}%
^{\alpha }\partial _{\alpha }=A^{-1/2}\partial _{u}$. $T^{\alpha \beta
}\beta _{\beta }$ has components along and orthogonal to $\hat{u}^{\alpha }$%
. We can thus write
\[
S^{\alpha }=S\hat{u}^{\alpha }+P^{\alpha }
\]
where $P^{\alpha }\hat{u}_{\alpha }=0$. The divergence of $S^{\alpha }$ is
\begin{equation}
S_{\,;\alpha }^{\alpha }=SA^{-1/2}[\dot{S}/S+A^{-1}(\dot{m}/r)]+P_{\,;\alpha
}^{\alpha }  \label{div-s}
\end{equation}
wherein the Vaidya luminosity gradient appears. The logarithm of $S$
measures the number of available states and we see here that $\dot{m}/r$
provides a measure of the rate of change in the number of states.

\section{Discussion}

Luminosity is one of the key observations made of astrophysical objects (a
mass-luminosity relation is an important feature of main sequence stars).
Use of the Bach current to compute luminosity gradients provides a covariant
method for obtaining an astrophysically important\ quantity. We have
identified the luminosity gradient with entropy production in the Vaidya
exterior. The Vaidya exterior dissipates energy by sending
''short-wavelength photons'' to null infinity. For an equilibrium ensemble
of star models, it is the Bach current and the rate of entropy increase that
allows use of the fluctuation-dissipation theorem to measure the rate at
which an initial deviation from equilibrium decays

A full development of dissipative thermodynamics using the Bach current
requires analyzing the collapsing Vaidya interior, which we leave for future
work. The Vaidya spacetime is only a first approximation to interesting
systems.

{\bf ACKNOWLEDGMENTS}

I'm indebted to Josh Goldberg, Jean Krisch, and Ted Newman for reading and
commenting on early versions of this work

\end{document}